\documentclass[prd,footinbib,aps,superscriptaddress,tightenlines,preprintnumbers,twocolumn]{revtex4}

\usepackage{epsfig,latexsym,amsmath,amssymb}
\usepackage[usenames]{color}
\newcommand\beq{\begin{equation}}
\newcommand\eeq{\end{equation}}

\newcommand\nn{\nonumber}

			\def\k{\kappa}

\def\k3{{\bar{k}_3}}

\def\OMIT#1{}
\begin{document}

\preprint{\vbox{\hbox{CALT-68-2684}\hbox{UCSD/PTH 08-03}}  } 

\title{Causality as an emergent  macroscopic phenomenon: The Lee-Wick O(N) model} 

\author{Benjam\'\i{}n Grinstein}
\email[]{bgrinstein@ucsd.edu}
\affiliation{Department of Physics, University of California at San Diego, La Jolla, CA 92093}

\author{Donal O'Connell}
\email[]{donal@ias.edu}
\affiliation{School of Natural Sciences, Institute for Advanced Study, Princeton, NJ 08540}

\author{Mark B. Wise}
\email[]{wise@theory.caltech.edu}
\affiliation{California Institute of Technology, Pasadena, CA 91125}

\begin{abstract}

In quantum mechanics the deterministic property of classical physics
is an emergent phenomenon appropriate only on macroscopic
scales. Lee and Wick introduced Lorentz invariant quantum theories
where causality is an emergent phenomenon appropriate for
macroscopic time scales. In this paper we analyze a Lee-Wick version
of the $ O(N)$ model.  We argue that in the large $N$ limit this
theory has a unitary and Lorentz invariant $S$ matrix and is
therefore free of paradoxes in scattering experiments. We discuss
some of its acausal properties.

\end{abstract}

\date\today
\maketitle


\section{Introduction}

It is interesting to try to understand if one or more of the pillars
of modern physics may be violated by a theory that gives approximately
the same experimental results as ordinary relativistic quantum field
theory for experiments that are presently accessible.  One such pillar
is causality.  Theories that are not causal appear, at first glance,
to be fraught with paradoxical behavior that renders them
inconsistent. In the late 1960's Lee and Wick~\cite{LW1},\cite{LW2} proposed an extension of
quantum electrodynamics where the Pauli-Villars regulator is treated
as a finite mass scale. In this theory the Fourier transform of
the gauge field two-point function has massive ``Lee-Wick photon''
poles with wrong sign residues. It is easy to show that this theory
is equivalent to a higher derivative theory. Naively such theories
are unstable and not unitary. Lee and Wick and Cutkoski, Landshoff,
Olive and Polkinghorne~\cite{CLOP}
gave rules  (the
  ``LW'' and ``CLOP prescriptions'') for calculating perturbative scattering amplitudes in this
higher derivative theory that, for a wide class of Feynman diagrams,
overcame these difficulties yielding Lorentz invariant and unitary
scattering amplitudes.   However Lee-Wick electrodynamics is not causal
for microscopic time scales. Their ideas provide a framework for
studying acausal theories where the acausality is only detectable in
experiments that can access very high energies and/or very short time
scales~\cite{Coleman}.

In recent papers we extended the work of Lee and Wick to non-Abelian
gauge theories and argued that they can solve the hierarchy
puzzle~\cite{us1,us2}. Even if the ideas of Lee and Wick are not
relevant for the hierarchy puzzle it is worth exploring the physics of
acausal theories and examining their consistency. Previous work has
some limitations. While the LW and CLOP prescriptions have been shown
to give Lorentz invariant scattering amplitudes in a large class of
Feynman graphs, it is not known whether this is true to all orders in
perturbation theory.
Moreover, there are serious obstacles to a non-perturbative path
integral formulation for the Lee-Wick theory with a CLOP
prescription~\cite{BG}. Other formulations with prescriptions
different from CLOP's can have a non-perturbative path-integral
formulation \cite{van Tonder:2006ye,Liu:1994zy}, but these have yet to
be shown to give a Lorentz invariant S matrix\cite{Nakanishi:1971kn}.
Perhaps there is some subtle obstacle that prevents the construction
of non-trivial Lee-Wick theories that have a unitary and Lorentz
invariant $S$ matrix to all orders in perturbation theory. We will
argue that this is not the case since in large $N$ the Lee-Wick $O(N)$
model provides an example of an acausal theory that has a unitary and
Lorentz invariant $S$ matrix.

The leading behavior at large $N$ of the scattering amplitudes in the
$O(N)$ scalar model can be calculated~\cite{ON}. In this paper we consider the
Lee-Wick version of this theory and argue that at large $N$ the
scattering matrix is unitary and Lorentz invariant. If the theory has
unitary time evolution there will be no paradoxes in experiments that
involve normal scalars in the prepared initial state and the observed
final state.  After all, for any initial state there are various
possible orthogonal final states and the probability for each of them
occurring sums to unity. This theory has unconventional acausal
effects, some of which we illustrate with explicit calculations, but
that does not mean it is inconsistent.

If a Lee-Wick extension of the standard model is relevant for the
hierarchy puzzle then the acausal effects can be studied indirectly in
high energy accelerator experiments through unusual interference
effects associated with a Lee-Wick resonance \cite{Rizzo}, like the wrong
sign of the phase shift of the resonant scattering amplitude.
However, there is no compelling reason that acausal effects should
occur at the weak scale; perhaps low energy supersymmetry or a warped
extra dimension provide the solution to the hierarchy puzzle. There are constraints on the 
masses of the Lee-Wick resonances from precision electroweak physics. These are quite strong
because integrating out the Lee-Wick resonances gives tree level contributions to the oblique parameters $S$ and $T$~\cite{Alvarez}.

It is possible that acausal effects, of the type we are studying in
this paper, arise from the extension of the standard model to include
a quantized theory of gravity.  String theory is an extension of the
standard model that includes a consistent quantum theory of
gravity. At the present time it is widely accepted that string theory
is realized in nature. This is because of a lack of alternatives and
because, even though string theory is highly constrained, it has
compactifications with enough light degrees of  freedom to accommodate
the known standard model physics as well as gravity. However, there is
no experimental evidence to support the hypothesis that string theory
is realized in nature. Therefore, even if string theory is
incompatible with the type of acausality we are studying, it seems
worth keeping an open mind on this issue and contemplating the
possibility that acausal effects occurring on time scales of order the
Planck time may arise from the extension of the standard model to
include quantum gravity
\footnote{Higher derivative gravity, a
    theory in which  terms quadratic in the curvature tensor are added
    to the usual Einstein-Hilbert action,  is
    known to be  renormalizabe and asymptotically free. Partly for the
    reasons that motivate this paper, it is not known whether this
    theory can be consistently formulated to all orders in
    perturbation theory.}. 

 In \cite{Wu} it was argued that gravitational radiative corrections
can induce higher derivative terms of the Lee-Wick type. 
In \cite{Tom} a Lee-Wick theory of gravity was considered.

In the auxiliary field formulation of the $O(N)$ model at large $N$
the only loop diagrams that enter the calculation of scattering
amplitudes is the one-loop 1PI auxiliary field self-energy.  In the
Lee-Wick extension of the $O(N)$ model this Feynman diagram can be
treated using the LW and CLOP prescriptions. Hence this theory has a
unitary and Lorentz invariant $S$ matrix at large $N$ and this toy
model provides a convenient laboratory to study the physics of acausal
theories.  In this paper we explore the Lee-Wick $O(N)$ model. We show
by explicit calculation that the two particle scattering amplitudes
satisfy the optical theorem, argue that the $S$ matrix is unitary and
calculate the acausal behavior that arises in some experiments.

Higher derivative versions of the $O(N)$ model at large $N$ have been
studied before \cite{Jansen:1992xv} and the question of
unitarity is taken up by Liu in Ref.~\cite{Liu:1994zy}. Our
investigations differ in several respects. The prescription that Liu
uses differs from CLOP's prescription. In Ref.~\cite{Liu:1994zy} terms
of order $\partial^6$ are added to the Lagrangian, arranged so that
{\it at tree level} there is a complex pair of poles in
propagators. The imaginary part of these poles is not associated with
a decay width. In our model only terms of order $\partial^4$ are added
to the Lagrangian so that at tree level the two poles in the
propagator are at real and positive values of $p^2$. It is
interactions that turn the wrong residue pole into a complete "di-pole,"
and the imaginary part is dictated by the physical width. Finally,
Liu's proof of unitarity is indirect\footnote{Liu does check
explicitly that the optical theorem holds for the scattering of
Goldstone bosons in the spontaneously broken version of his theory.}
while we demonstrate unitarity by direct calculation.

\section{Time Dependence in a Scattering Experiment}

We start our discussion by entertaining the question, how does one
 go about testing causality or looking for causality violation in a
 theory that gives only an $S$ matrix? One may wonder if this is
 possible at all since formally the $S$ matrix relates states of the infinite
 past to those in the infinite future. Intuitively it is clear that
 this is no impediment: we may prepare two localized wave packets in the infinite
 past to travel toward each other and set up detectors to look for the
 outcome of their collision. Moreover, if both the distances traveled by
 the prepared wave-packets to the
 collision point and the distances from this point to the detectors
 are truly macroscopic, then there is an $S$ matrix description of this
 process. Clearly, information on the position and timing of first
 detection of collision products can then give information on the
 causal behavior of the interaction. 

This section formalizes these statements mathematically. In theories
with normal causal behavior, a resonant collision that takes place at
some space-time point, $z_0$, results in the production of outgoing stable
particles that appear to arise from a second space-time point, $z_0'$. This second point
occurs later in time ($t_0'>t_0$) and is typically separated  from the
collision point by proper time of the order of $1/\Gamma$,  the inverse of the
width of the resonance. The distribution is a decaying exponential.  This is encapsulated in
Eq.~\eqref{eq:normtreetime}, in which the separation between the two
points is $w=z_0'-z_0$.

The situation is quite different for resonant collisions through 
a  Lee-Wick resonance. Here the detected particles appear to come from
$z_0'$ which occurs earlier than $z_0$ in time ($t_0'<t_0$). The
points are still separated by proper time of the order of $1/\Gamma$,
still distributed as an  exponential that decays away from $z_0$. This
is the content of the final equation in this chapter,
Eq.~\eqref{eq:normtreetime1}. 

We have been careful to state that the collision
  ``appears to'' take place at $z_0'$. The measurement is made a long
  time after and a long
  distance from the collision region. Within the quantum mechanical
  $S$ matrix formalism there is no means by which we can investigate,
  nor is there meaning to, the question of precisely where or when the
  collision takes place. This observation is important in
  understanding the interpretation of the results in the case where
  the collision goes through a Lee-Wick resonance, where normal causal
  behavior is violated.

\subsection{Kinematics}
\label{sec:kinematics}

We prepare from stable particles of mass $m$ an initial state
consisting of two wave packets traveling towards each other from far
away. They are initially localized about spacetime points $y_1$ and
$y_2$ which we can assume are space-like separated, $(y_2-y_1)^2<0$
and have large negative time components (we imagine the interaction
will take place at around zero time). So we take $y_i^0<0$ and
$|y_i^0|\gg 1/m$ where $m$ is the mass of the particles. Since they
will have to travel a long distance to the interaction point we also
take $|\vec y_i|\gg1/m$.\footnote{In fact we will need all
      inverse masses {\it and momenta} to be small compared to all the
      separations involved in the problem, including eventually the
      distance between collision and decay points, $z_0'-z_0$. By
      restoring Planck's constant, $|\vec y_i|\gg\hbar/m$, we see that
      a simpler, equivalent assumption is that we analyze the
      kinematics in the semiclassical limit. We emphasize, however,
      that the exact scattering amplitude is used in this section,
      that is, no approximation for the scattering need be made in
      discussing time dependence in a scattering experiment. }

We also want the wave packets to have specific momenta. That is, their
Fourier transforms are localized about $p_1=mv_1$ and $p_2=mv_2$. Of
course, the momenta have to point towards each other so that there is
a collision. The collision occurs at a point $z_0$ such that
\begin{equation}
\label{z0}
\frac{z_0-y_1}{\tau_1}=v_1\quad\text{and}\quad
\frac{z_0-y_2}{\tau_2}=v_2
\end{equation}
where $\tau_i=\sqrt{(z_0-y_i)^2}$ is the proper time along the world
line of the particle from the start point to the interaction point.

So we take for the initial state
\begin{equation}
|\psi_{{\rm in}}\rangle= \int d^4x_1\,d^4x_2\,
f_1(x_1-y_1)f_2(x_2-y_2)\phi(x_1)\phi(x_2)|0\rangle
\end{equation}
with $f_i(x)$ concentrated about $x=0$, and the Fourier transform
\begin{equation} \tilde f_i(k_i)=\int d^4x e^{ik_i\cdot x}f_i(x)
\end{equation}
concentrated about $k_i=p_i$ with $p_i^2=m^2$. Here $\phi(x)$ is a
real scalar field that, when acting on the vacuum, creates a stable
particle of mass $m$.

Similarly, we will set up two detectors for the outgoing particles
that each record only at a particular point in space at a specific
time, that is, at spacetime points $y_i'$. These points can also be
taken as space-like separated and at late times and large distances,
$y_i^{\prime0}\gg 1/m$ and $|\vec y'_i|\gg1/m$. And we want to absorb
specific momenta, $p_i'=mv_i'$. If the outgoing particles emerge from
a point $z_0'$ then 
\begin{equation}
\frac{y_1'-z_0'}{\tau'_1}=v_1'\quad\text{and}\quad
\frac{y_2'-z_0'}{\tau'_2}=v_2'.
\end{equation}

 So we take for the final state 
\begin{equation}
 |\psi_{{\rm out}}\rangle= \int d^4x'_1\,d^4x'_2\,
 g_1(x'_1-y'_1)g_2(x'_2-y'_2)\phi(x'_1)\phi(x'_2)|0\rangle 
\end{equation} 
with $g_i(x)$ concentrated about $x=0$, and their Fourier transforms
concentrated at $p_i'=m v_i'$.

Consider now the amplitude for the state $ |\psi_{{\rm in}}\rangle$ to evolve into the state $
|\psi_{{\rm out}}\rangle$, 
\begin{widetext}
\begin{equation}
\langle\psi_{{\rm out}}|\psi_{{\rm in}}\rangle= 
\int d^4x_1\,d^4x_2\, d^4x'_1\,d^4x'_2\,
g_1^*(x'_1-y'_1)g_2^*(x'_2-y'_2)
f_1(x_1-y_1)f_2(x_2-y_2)\langle0|\phi(x'_2)\phi(x'_1)\phi(x_1)\phi(x_2)|0\rangle.
\end{equation}
Since we have initial points that are space-like separated the order of
the fields at $x_1$ and $x_2$ is irrelevant and the same goes for the
fields at  $x_1'$ and $x_2'$. Also the fields at $x_i'$ have later
times than those at $x_i$. So we can replace  the product of fields above by the
time ordered product, which is just the 4-point Green function, 
\begin{multline}
\langle0|T[\phi(x'_2)\phi(x'_1)\phi(x_1)\phi(x_2)]|0\rangle
=
\int \prod_{i}\frac{d^4k_i}{(2\pi)^4} \frac{d^4k_i^{\prime}}{(2\pi)^4}e^{i( k_1\cdot x_1+ k_2\cdot x_2- k_1^{\prime}\cdot x_1^{\prime}- k_1^{\prime}\cdot x_1^{\prime})}(2\pi)^4\delta^{(4)}( k_1+k_2-k_1^{\prime}-k_2^{\prime})  \\
 \times  \prod_i\frac{i}{k_i^2-m^2+i\epsilon}\frac{i}{k_i^{\prime 2}-m^2+i\epsilon}\Gamma^{(4)}(k_1,k_2,-k_1^{\prime},-k_2^{\prime}).
\end{multline}
We have written this in Fourier space in terms of the amputated
4-point function.  We will consider cases where the amputated 4-point
function $\Gamma^{(4)}(k_1,k_2,-k_i^{\prime},-k_2^{\prime})$ is
the sum of three terms which depend respectively on the Mandelstam
variables, $s$, $t$ and $u$. At large separation $|z_0-z_0'|$ the
  amplitude $\langle\psi_{{\rm out}}|\psi_{{\rm in}}\rangle$ will be
  negligible except when there is a narrow $s$-channel resonance in
  the 4-point function. This is clear if we use customary causal intuition, that a narrow
  resonance can be thought of as a long lived unstable particle
  produced at $z_0$ decaying {\it later} at $z_0'$, but
  mathematically it is true even when the resonance ``decays'' at
  $z_0'$ {\it before} $z_0$. Therefore to examine the leading dependence of
  $\langle\psi_{{\rm out}}|\psi_{{\rm in}}\rangle$ on $z_0-z_0'$ , for
  large $|z_0-z_0'|$, only the term that depends on $s=(k_1+k_2)^2$ is
  important and we drop the other pieces. We denote this term
  by   $\Gamma^{(4)}_s(s)$.

Now, multiplying the above by
\begin{equation}
1=\int \frac{d^4q}{(2\pi)^4}(2\pi)^4\delta(k_1+k_2-q)
\end{equation}
 we get
\begin{align}
\label{FxGgiven}
\langle\psi_{{\rm out}}|\psi_{{\rm in}}\rangle&=\int d^4x_1\,d^4x_2\, d^4x'_1\,d^4x'_2\,
g_1^*(x'_1-y'_1)g_2^*(x'_2-y'_2)
f_1(x_1-y_1)f_2(x_2-y_2)\nn\\
&\qquad\times\int
\prod_{i}\frac{d^4k_i}{(2\pi)^4}\frac{d^4k'_i}{(2\pi)^4}
\frac{d^4q}{(2\pi)^4}(2\pi)^4\delta^{(4)}(k_1+k_2-q) (2\pi)^4\delta^{(4)}(k_1'+k_2'-q) 
\nn\\ &\qquad\qquad\qquad\times
e^{i(k_1\cdot x_1+k_2\cdot x_2-k_1'\cdot x_1'-k_2'\cdot x_2')}
\prod_{i}\frac{i}{k_i^2-m^2+i\epsilon}\frac{i}{k_i^{\prime 2}-m^2+i\epsilon}\Gamma_s^{(4)}(q^2)\nn\\
&=\int \frac{d^4q}{(2\pi)^4} \tilde
F(q)\tilde G(q) \Gamma_s^{(4)}(q^2)
\end{align}
where we have introduced
\begin{equation}
\label{F-given}
\tilde F(q)=\int
\prod_{i}\frac{d^4k_i}{(2\pi)^4} (2\pi)^4\delta(k_1+k_2-q)\, e^{i\sum
  k_i\cdot y_i}\tilde f_1(k_1)\tilde f_2(k_2)\prod_i\frac{i}{k_i^2-m^2+i\epsilon}
\end{equation}
and
\begin{equation}
\tilde G(q)=\int
\prod_{i}\frac{d^4k_i'}{(2\pi)^4} (2\pi)^4\delta(k'_1+k'_2-q)\, e^{-i\sum
  k_i'\cdot y_i'}\tilde g^*_1(k_1)\tilde g^*_2(k_2)\prod_i\frac{i}{k_i^{\prime2}-m^2+i\epsilon}~.
\end{equation}

The integral in \eqref{F-given} can be broken into two single particle
integrals by representing the delta function as an integral,
\begin{equation}
\label{F-given2}
\tilde F(q)=\int d^4z\int
\prod_{i}\frac{d^4k_i}{(2\pi)^4} \, e^{iz\cdot(q-k_1-k_2)+i\sum
  k_i\cdot y_i}\tilde f_1(k_1)\tilde
f_2(k_2)\prod_i\frac{i}{k_i^2-m^2+i\epsilon}
=\int d^4z\, e^{iz\cdot q}I_1(z)I_2(z)
\end{equation}
\end{widetext}
where
\begin{equation}
I_i(z)=\int \frac{d^4k_i}{(2\pi)^4} \tilde f_i(k_i) e^{i k_i\cdot
  (y_i-z)}\frac{i}{k_i^2-m^2+i\epsilon}\;.
\end{equation}

We now estimate $I_i$. To this end notice that  each
component of $y_i-z_0$ is much larger in magnitude than $1/m$, by assumption. We
will compute for $z\approx z_0$ and come back later to see that the
integral in the last line of \eqref{F-given2} has support localized
around $z= z_0$. Rewrite $I_i$ by exponentiating the propagator,
using the $i\epsilon$ of the normal particle propagator
\begin{equation}
I_i=\int_0^\infty ds  \int \frac{d^4k_i}{(2\pi)^4} \tilde f_i(k_i) e^{i k_i\cdot
  (y_i-z)} e^{is(k_i^2-m^2+i\epsilon)}.
\end{equation}

We do first the $k_i$ integration by the method of stationary
phase.\footnote{As explained above, this is justified by
      considering $\hbar\to0$.}  The condition that the phase be
stationary is
 \begin{equation} \frac{\partial}{\partial k^\mu_i} \left( k_i\cdot
  (y_i-z) + s k_i^2\right)=0, 
\end{equation} 
which implies that
\begin{equation}
k_i=\frac{z-y_i}{2s}.  
\end{equation} 
So we get
\begin{widetext}
\begin{equation}
I_i\simeq \frac1{(2\pi)^4}
e^{-i\frac{\pi}2}\int_0^\infty ds \left(\sqrt{\frac{\pi}{s}}\right)^4
\tilde f_i\left(\frac{z-y_i}{2s}\right) 
e^{- i\frac{(z-y_i)^2}{4s}+ is(-m^2+i\epsilon)}.
\end{equation}
Now do the $s$-integration also using stationary phase. The condition
is
\begin{equation}
\frac{\partial}{\partial
  s}\left[\frac{(z-y_i)^2}{4s}-s(-m^2+i\epsilon)\right]_{s=s_i}=0
\end{equation}
which implies that
\begin{equation}
s_i=\frac{\sqrt{(z-y_i)^2}}{2m}
\end{equation}
and leads to 
\begin{equation}
I_i \simeq  \frac{e^{-i\frac{3\pi}4}}{(2\pi)^4}\frac{\pi^{5/2}}{ms_i^{3/2}}
\tilde f_i\left(\frac{z-y_i}{2s_i}\right) e^{-im\sqrt{(z-y_i)^2}}.
\end{equation}

We now put the pieces together. First, putting the result for $I_i$ above into  $\tilde F$ as given in \eqref{F-given2} we have
\begin{equation}
\tilde F\simeq \int d^4z e^{i z\cdot q}\frac{-i}{2^8\pi^3m^2(s_1s_2)^{3/2}}
\tilde f_1\left(\frac{z-y_1}{2s_1}\right) 
\tilde f_2\left(\frac{z-y_2}{2s_2}\right) 
e^{-im\sqrt{(z-y_1)^2}-im\sqrt{(z-y_2)^2}}.
\end{equation}
\end{widetext}
Let's investigate this function. Using stationary phase, which is
justified because we can think of $q$ or $m$ as large, or
  equivalently,  because we can take $\hbar\to0$, we see that the
stationary phase condition gives $z=z_*$ with $z_*$ satisfying
\begin{equation}
q-m\frac{z_*-y_1}{\sqrt{(z_*-y_1)^2}}-m\frac{z_*-y_2}{\sqrt{(z_*-y_2)^2}}=0\,.
\end{equation}
The second and third terms are just the arguments of the functions
$\tilde f_i$, which, we recall, have localized support. So the
function $\tilde F$ is non-zero only for
\begin{equation}
\frac{z_*-y_1}{\sqrt{(z_*-y_1)^2}}=v_1\quad\text{and}\quad\frac{z_*-y_2}{\sqrt{(z_*-y_2)^2}}=v_2
\end{equation}
which implies the support is at 
\begin{equation}
q=mv_1+mv_2=p_1+p_2\,.
\end{equation}
Moreover, we know how to solve for $z_*$ in the region where the
integral has support: by Eq.~\eqref{z0} we see that $z_*=z_0$, the
point where the interaction takes place. We conclude then that
\begin{equation}
\tilde F(q)\simeq e^{iq\cdot z_0}\hat F(q)
\end{equation}
where $\hat F$ is smooth\footnote{Only the exponential has a
      fast oscillation with the variable $q$ as $\hbar\to0$; the
      remaining factor does contain an exponential involving
      $1/\hbar$, namely $\exp(-im(\tau_1+\tau_2)/\hbar)$, however this
      is $q$-independent.} and has support localized at $q=p_1+p_2$.
Note that we are absorbing the phase factor, $\exp
(-im\sqrt{(z_0-y_1)^2})\exp (-im\sqrt{(z_0-y_2)^2})= \exp(-im\tau_1-im
\tau_2)$, 
into the definition of $\hat F$.  Similarly, we conclude
that
\begin{equation}
\tilde G(q)\simeq e^{-iq\cdot z'_0}\hat G(q)
\end{equation}
with $\hat G$ smooth and with
 support localized at $q=p'_1+p'_2$. 

Using this information on the structure of $\tilde F$ and $\tilde G$ in
\eqref{FxGgiven} we have
\begin{equation}
\label{keyresult}
\langle\psi_{{\rm out}}|\psi_{{\rm in}}\rangle
\simeq \int \frac{d^4q}{(2\pi)^4} e^{-iq\cdot(z_0'-z_0)}\hat
F(q)\hat G(q) \Gamma_s^{(4)}(q^2)\,.
\end{equation}
Eventually we will calculate the $w=z_0'-z_0$ dependence of
$\langle\psi_{{\rm out}}|\psi_{{\rm in}}\rangle$ for large $w^0$ in
the Lee-Wick $O(N)$ model. Before considering the $O(N)$ model its
instructive to consider some simpler examples.

\subsection{Standard resonant behavior}
Here we recover the more familiar time dependence associated with
resonant $s$-channel exchange in a toy model with two real scalar
fields $\phi$ and $\chi$ and Lagrange density,
\begin{equation}
{\cal L}={1 \over 2}\partial_{\mu} \phi \partial^{\mu}\phi -{1 \over 2}m^2\phi^2 +{1 \over 2}\partial_{\mu} \chi \partial^{\mu}\chi -{1 \over 2}M^2\chi^2 +{g \over 2}\phi^2\chi.
\end{equation}
For simplicity we work at weak coupling $g/M \ll 1$ and also take $m/M
\ll 1$. 
The Fourier transform of $\chi$'s two-point function has the
form,
\begin{equation}
\label{propchi}
D_{\chi}(p^2)={i \over \pi} \int_{4m^2}^\infty  { d}s {\rho(s) \over  p^2-s+i \epsilon} \simeq {i \over p^2-M^2+iM\Gamma}
\end{equation}
where at order $g^2$ in perturbation theory the $\chi$ width is equal
to $\Gamma= g^2/(32 \pi M)$. In the narrow resonance approximation
\begin{equation}
\label{narrowres}
\rho(s) \simeq {M\Gamma \over (s-M^2)^2+M^2\Gamma^2}\;.
\end{equation}
Since $\rho(s)$ in Eq.~(\ref{narrowres}) is strongly peaked near
$s=M^2$ we can extend the $s$-integration in Eq.~(\ref{propchi}) over
the whole real $s$ line. Performing this integration using contour
methods with the value for $\rho(s)$ in Eq.~(\ref{narrowres})
reproduces the usual resonance form of the propagator on the far right
hand side of Eq.~(\ref{propchi}).

We calculate the dependence of $\langle\psi_{{\rm out}}|\psi_{{\rm
    in}}\rangle$ on $w=z_0'-z_0$ for large proper time $\sqrt{w^2}$,
under the assumption that the functions $F(q)$ and $G(q)$ are slowly
varying and have most of their support around $q=p_1+p_2$ and
$q=p_1'+p_2'$, respectively.  We also assume that the coupling $g$ is
small.  The amputated four point function is,
\begin{equation}
\Gamma_s^{(4)}(q^2)=-g^2D_{\chi}(q^2),
\end{equation}
 and so Eq.~(\ref{keyresult}) becomes,
 \begin{widetext}
 \begin{equation}
\langle\psi_{{\rm out}}|\psi_{{\rm in}}\rangle=-i g^2\int {d^4q \over (2\pi)^4} {\hat F}(q){\hat G}(q) e^{-iqw}{1 \over q^2-M^2+iM \Gamma}=-g^2\int _0^{\infty}ds \int {d^4q \over (2\pi)^4}  {\hat F}(q){\hat G}(q)e^{-iqw}e^{is(q^2-M^2+iM\Gamma)}\,.
 \end{equation} 
 The integration over the components of the momentum $q$ is done using
 the stationary phase approximation. The stationary point is at
 $q=w/(2 s)$ and we find that (up to a constant phase),
  \begin{equation}
\langle\psi_{{\rm out}}|\psi_{{\rm in}}\rangle \simeq  {g^2 \over (2\pi)^2}\int _0^{\infty} ds \left({1 \over  2 s} \right)^2 {\hat F}(w/(2s)){\hat G}(w/(2s)) e^{-i(w^2/(4s)+sM^2)}e^{-\Gamma M s}.
 \end{equation} 
 Next we perform the $s$ integration using the stationary phase
 approximation. The stationary point is at ${s}= \sqrt{ w^2}/(2M)$ and
 we arrive at,
 \begin{equation}
\langle\psi_{{\rm out}}|\psi_{{\rm in}}\rangle \simeq {g^2 \sqrt{M} \over 2(2\pi \sqrt{ w^2})^{3/2} }{\hat F}(Mw/\sqrt{w^2}){\hat G}(Mw/\sqrt{w^2})e^{-iM\sqrt{ w^2}}e^{-\Gamma \sqrt{ w^2}/2}\;.
 \end{equation}
 Since the functions ${\hat F }(q)$ and ${\hat G}(q)$ are peaked
 around  $q=p_1+p_2$ and $q=p_1'+p_2'$, respectively, the
 amplitude is appreciable only if $p_1+p_2\approx p_1'+p_2'\approx
 Mw/\sqrt{w^2}$. In the center of mass frame,  the  initial state must be
prepared with total energy near $M$ 
and the detectors in the final state are
designed to find ordinary particles that are back to back with total energy near $M$. The amplitude is dominated by $w^0>0$
 and $ {\vec w} \simeq 0 $ and we can write, in the CM frame, 
 \begin{equation}
 \label{eq:normtreetime}
\langle\psi_{{\rm out}}|\psi_{{\rm in}}\rangle \simeq \theta(w^0){g^2 \sqrt{M} \over 2(2\pi w^0)^{3/2} }{\hat F}(Mw/\sqrt{w^2}){\hat G}(Mw/\sqrt{w^2})e^{-iMw^0}e^{-\Gamma w^0/2}\;.
 \end{equation}
 Note that theta function means that the out-going $\phi$ wave-packets
 appear to emerge from the $\chi$ decay at a time $z_0'$ that is after
 the time $z_0$ that the incoming $\phi$ wave packets collide. The
 factor of ${\rm exp}(-\Gamma w^0/2)$ gives the characteristic
 exponential decay associated with the $\chi$ resonance and the factor
 of $(1/w^0)^{3/2}$ arises from the spreading of the $\chi$
 wave packet.
 \end{widetext}
 
 \subsection{Lee-Wick resonant behavior}
\label{sec:LWresBeh} 
Here we illustrate the acausal behavior of $\langle\psi_{{\rm
    out}}|\psi_{{\rm in}}\rangle$ in the simple Lee-Wick toy-model
introduced in \cite{us1}. The Lagrange density for this theory is,
 \begin{equation}
 {\cal L}={1 \over 2}\partial_{\mu}{\hat \phi}\partial^{\mu} {\hat \phi} -{1 \over 2M^2}(\partial^2 {\hat \phi})^2 -{1 \over 2}m^2 {\hat \phi}^2 - {1 \over 3!}g{\hat \phi}^3.
 \end{equation}
 The higher derivative term can be removed by adding a field ${\tilde
   \phi}$ in terms of which the Lagrange density becomes,
 \begin{equation}
 {\cal L}= {1 \over 2}\partial_{\mu}{\hat \phi}\partial^{\mu} {\hat \phi} -{1 \over 2}m^2 {\hat \phi}^2 -{\tilde \phi} \partial^2{\hat \phi}+{1 \over 2}M^2{\tilde \phi}^2-{1 \over 3!}g{\hat \phi}^3.
 \end{equation}
 Next we define $\phi=\hat \phi+\tilde \phi$ since in terms of $\phi $
 and ${\tilde \phi}$ the two derivative terms are not coupled. The
 Lagrange density now takes the form,
 \begin{multline}
 {\cal L}={1 \over 2}\partial_{\mu}\phi\partial^{\mu}\phi -{1 \over 2}\partial_{\mu}{\tilde \phi}\partial^{\mu}{\tilde \phi}+{1 \over 2}M^2{\tilde \phi}^2-{1 \over 2}m^2(\phi-\tilde \phi)^2 \\
  -{1 \over 3!}g(\phi-\tilde \phi)^3.
 \end{multline}
 Provided that $M>2m$ the mass matrix can be diagonalized  by a symplectic transformation 
 \begin{subequations}
 \label{symplectic}
 \begin{eqnarray}
 \phi &=& \cosh \theta \phi' +{\sinh}\theta {\tilde \phi}', \\
 {\tilde \phi}&=&{\sinh}\theta \phi' +{\cosh}\theta {\tilde \phi}'
 \end{eqnarray}
 where
 \begin{equation}
 {\tanh}2\theta=-2m^2/(M^2-2m^2).
 \end{equation}
 \end{subequations}
The Lagrange density then takes the form
 \begin{eqnarray}
 {\cal L}&=&{1 \over 2}\partial_{\mu}\phi' \partial^{\mu}\phi' -{1 \over 2}{m'}^2{\phi'}^2-{1 \over 2}\partial_{\mu}{\tilde \phi'}\partial^{\mu}{\tilde \phi'}+{1 \over 2}{M'}^2{{\tilde {\phi'}}}^2 \nonumber \\
  &-&{1 \over 3!}g({\rm cosh} \theta -{\rm sinh}\theta)^3(\phi'-\tilde \phi')^3.
 \end{eqnarray}
 Defining $g'=g({\rm cosh} \theta -{\rm sinh}\theta)^3$ and then
 dropping all the primes gives the Lagrange density in a convenient
 form.  For simplicity we take $m \ll M$.
 
 The free field propagator for the normal scalar takes the usual form
 $i/(p^2-m^2)$, however, the free field propagator for the Lee-Wick
 field, $\tilde \phi$, is $-i/(p^2-M^2)$ which differs by an overall
 minus sign from a conventional scalar of mass $M$. That minus sign
 means that the propagator that one gets from summing Lee-Wick self
 energy insertions develops a complex pole at
 $p^2=M_c^2=M^2+iM\Gamma$. Note that this is has positive imaginary
 part. Since the propagator remains real and regular on a segment of
 the real axis (below the two normal particle cut), the propagator
 satisfies Green's reflection principle, $(D_{\tilde \phi}
 (p^2))^*=D_{\tilde \phi} (p^{2\star})$. There is therefore a second
 pole at $p^2={ M_c^{*2}}$. The propagator can be written as the sum
 of terms with poles at $M_c^2=M^2+iM\Gamma$, $M_c^{*2}$ and the two
 particle cut,
 \begin{equation}
\label{fullLWprop}
  D_{\tilde \phi} (p^2)={-i \over p^2 - M_c^2}+ {-i \over p^2-{M_c^{*2} }}
  +{i \over \pi} \int_{4m^2}^\infty  { d}s {\rho(s) \over  p^2-s+i \epsilon}.
 \end{equation}
These poles  must not give rise to
additional imaginary parts in matrix elements since only the normal
$\phi$ particle is in the spectrum of the theory.

In the narrow resonance approximation the spectral density,
$\rho(s)$, is again given by Eq.~(\ref{narrowres}).  The spectral
representation for the propagator given in Eq.~\eqref{propchi}
contains no poles in $p^2$ but rather has a cut associated with the
integral over $s$. However, for a Lee-Wick resonance there are
poles at $p^2=M_c^2$ and $p^2={ M_c^{*2}}$. In the narrow resonance
approximation the term in Eq.~\eqref{fullLWprop} that has a
pole at $p^2={M_c^*}^2$ cancels against the term that contains the
integral over $s$ ({\it i.e.} the cut piece),
 \begin{equation}
 \label{leewickresprop}
 D_{\tilde \phi} (p^2)\simeq{-i \over p^2-M^2 -iM\Gamma}~,
 \end{equation}
 where $\Gamma \simeq  g^2/(32 \pi M)$.

We want to calculate the large $w^0$ behavior of $\langle\psi_{{\rm
    out}}|\psi_{{\rm in}}\rangle$ that arises from $s$-channel exchange
of the Lee-Wick resonance $\tilde \phi$ at tree level making the same
assumptions that we did in the case where there was $s$-channel
exchange of the ordinary resonance $\chi$. In the case of Lee-Wick
resonant exchange,
\begin{equation}
\Gamma_s^{(4)}(q^2)=-g^2D_{\tilde \phi}(q^2),
\end{equation}
with $D_{\tilde \phi}$ given in Eq.~(\ref{leewickresprop}).  We follow
the same steps that were used for the $\chi$ case. Since the width
term in the propagator has the opposite sign  the phase of the
exponential proportional to $s$ must be flipped to get
convergence at infinity. Hence, we find that,
\begin{widetext}
 \begin{equation}
\langle\psi_{{\rm out}}|\psi_{{\rm in}}\rangle =ig^2\int {d^4q \over (2\pi)^4} {\hat F}(q){\hat G}(q) e^{-iqw}{1 \over q^2-M^2-iM \Gamma}=-g^2\int _0^{\infty}ds \int {d^4q \over (2\pi)^4}  {\hat F}(q){\hat G}(q)e^{-iqw}e^{-is(q^2-M^2-iM\Gamma)}.
\end{equation} 
The stationary point for the $q$ integration is now at,  $q=-w/(2 s)$ and we find that  (up to a constant phase),
 \begin{equation}
\langle\psi_{{\rm out}}|\psi_{{\rm in}}\rangle \simeq  {g^2 \over (2\pi)^2}\int _0^{\infty} ds \left({1 \over  2 s} \right)^2 {\hat F}(-w/(2s)){\hat G}(-w/(2s)) e^{i(w^2/(4s)+sM^2)}e^{-\Gamma M s}.
 \end{equation} 
The stationary point for the $s$ integration is in the same place as before, ${s}=  \sqrt{ w^2}/(2M)$  and we arrive at, 
 \begin{equation}
   \langle\psi_{{\rm out}}|\psi_{{\rm in}}\rangle \simeq {g^2 \sqrt{M} \over 2 (2\pi \sqrt{ w^2})^{3/2} }{\hat F}(-Mw/\sqrt{w^2}){\hat G}(-Mw/\sqrt{w^2})e^{iM\sqrt{ w^2}}e^{-\Gamma \sqrt{ w^2}/2}.
 \end{equation}
 Since in the CM frame the functions ${\hat F }(q)$ and ${\hat G}(q)$
 are peaked around $q^0=M$ and ${\vec q}=0$ the amplitude is dominated
 by $w^0<0$ and $ {\vec w} \simeq 0 $ and we can write
 \begin{equation}
 \label{eq:normtreetime1}
\langle\psi_{{\rm out}}|\psi_{{\rm in}}\rangle \simeq \theta(-w^0){g^2 \sqrt{M} \over 2(2\pi |w^0|)^{3/2} }{\hat F}(-Mw/\sqrt{w^2}){\hat G}(-Mw/\sqrt{w^2})e^{-iMw^0}e^{\Gamma w^0/2}.
 \end{equation}

 The theta function in Eq.~(\ref{eq:normtreetime1}) means that the
 out-going $\phi$ wave-packets appear to emerge from the ${\tilde
   \phi}$ decay at a time $z_0'$ that is before the time $z_0$ that
 the incoming $\phi$ wave packets collide. The factor of ${\rm exp}(
 \Gamma w^0/2)$ gives backwards in time exponential decay of the
 ${\tilde \phi}$ resonance and the factor of $(1/|w^0|)^{3/2}$ arises
 from the spreading of the $\tilde \phi$ wave packet.

\end{widetext}

\section{Review of the $O(N)$ Model}

The theory contains $N$ scalar fields $\phi^a(x)$, $a=1, \ldots N$, and is invariant under orthogonal transformations of them, $\phi^a(x) \rightarrow  \phi'^a(x)= O^{ab}\phi^b$. The Lagrange density is
\begin{equation}
\label{lag1}
{\cal L}={1 \over 2}\partial_{\mu}\phi^a \partial^{\mu}\phi^a-{ 1 \over 2} m_0^2 \phi^a \phi^a -{1 \over 8} \lambda_0 \left( \phi^a \phi^a \right)^2.
\end{equation} 
To get a sensible large $N$ limit of this theory we must take
$\lambda_0 \sim {\cal O }(1/N)$.  Then the $\phi \phi \rightarrow \phi
\phi$ scattering amplitude is $ {\cal O }(1/N)$, the $\phi \phi
\rightarrow \phi \phi \phi \phi$ amplitude is $ {\cal O }(1/N^2)$ {\it
  etc.}.  This is similar to the behavior of ${ QCD}$ in the large
number of colors limit \cite{'t Hooft1}. There color singlet mesons
$M$ with interpolating fields of the form\footnote{The flavor
  structure is suppressed here.} ${\bar q} q/{\sqrt N_c}$ have $MM
\rightarrow MM$ scattering amplitudes that are ${\cal O}(1/N_c)$ and
$MM \rightarrow MMMM$ scattering amplitudes that are ${\cal
  O}(1/N_c^2)$. However in the case of 
the $O(N)$ model the $\phi \phi
\rightarrow \phi \phi$ cross section averaged over initial values of
the $O(N)$ quantum number $a$ and summed over final values is ${\cal O
}(1/N)$ while in the large $N_c$ limit of $QCD$ $MM \rightarrow MM$
scattering cross sections are ${\cal O}(1/N_c^2)$.

  It is convenient to remove the quartic interaction term in
  Eq.~(\ref{lag1}) by introducing a non dynamical scalar $\sigma$, and
  make the $N$ dependence explicit by introducing
  $\lambda=N\lambda_0$.  The Lagrangian density takes the form,
\begin{equation}
\label{lag2}
{\cal L}={1 \over 2}\partial_{\mu}\phi^a \partial^{\mu}\phi^a -{ 1 \over 2} m_0^2 \phi^a \phi^a+{N \over  2\lambda} \sigma ^2 -{1 \over 2} \sigma \phi^a \phi^a.
\end{equation}
One can show that the Lagrange density in Eq.~(\ref{lag2}) is
equivalent to that in Eq.~(\ref{lag1}) by integrating out the scalar
$\sigma$ using its equations of motion, $\sigma =\lambda\phi^a
\phi^a/(2N)$. In this formulation the only interaction vertex is
between a sigma and two $\phi$'s.

It is straightforward to argue that in this formulation of the theory,
at large $N$, the only loop diagrams that must be computed is a 1-loop
$\sigma$ self energy, $\Sigma_0$ and a $\sigma$ tadpole. The physical
effects of the tadpole can be absorbed into the $\phi$ mass by making
the replacement, $m_0^2 \rightarrow m^2$. Using dimensional
regularization with $ \overline {MS}$ subtraction,
\begin{equation}
\Sigma_0(p^2)= - {N \over 32 \pi^2}\int_0^1 dx \log \left({m^2 - p^2x(1-x) - i\epsilon \over \mu^2}\right),
\end{equation}
where $\mu$ is the subtraction point. The sigma propagator is,
\begin{equation}
D_{\sigma}(p^2)={ i \over {1\over \lambda_0}}+{ i \over {1\over \lambda_0}} i \Sigma_0(p^2){ i \over {1\over \lambda_0}}+\ldots
={i \over {1/\lambda_0+\Sigma_0(p^2)}}.
\end{equation}
Writing the scattering matrix as $S=1+iT$, we define the scattering amplitude, ${\cal M} $, by
\begin{multline}
\langle k^\prime_1, c; k^\prime_2, d | T | k_1, a; k_2, b \rangle \\
=(2 \pi)^4 \delta^4(k_1+k_2-k_1^\prime-k_2^\prime) \mathcal{M}(k_1, a; k_2, b \rightarrow k_1^\prime, c; k_2^\prime, d)  .
\end{multline}
Using our expression for the sigma propagator, we find that the scattering amplitude is given by
\begin{equation}
\label{eq:scatt}
{\cal M}(k_1, a; k_2, b \rightarrow k_1^\prime, c; k_2^\prime, d)=-{\lambda \over N}\left(  {\delta_{ab}\delta_{cd} \over 1+\lambda \Sigma(s)}+ \ldots \right),
\end{equation}
where $s,t,u$ are the usual Mandelstam variables and the ellipses
denote the two terms similar to the one presented that are functions
of $t$ and $u$. Note we have written $\Sigma_0=N\Sigma$ to make all
the $N$ dependence explicit.

\subsection{Unitarity of the Two Particle Scattering Amplitudes}

With these results in hand, we can explicitly check unitarity of two
particle scattering in the $O(N)$ model to leading order in $1/N$ and
to all orders in $\lambda$. Unitarity of the $S$ matrix, i.e.,
$S^\dagger S = 1$, is equivalent to $i (T^\dagger - T) = T^\dagger
T$. Taking the two particle matrix element of this equation gives
\begin{widetext}
\begin{equation}
i \left(\mathcal{M}(k_1^\prime, c; k_2^\prime, d \rightarrow k_1, a; k_2, b)^* - \mathcal{M}(k_1, a; k_2, b \rightarrow k_1^\prime, c; k_2^\prime, d) \right) = \sum_\psi \mathcal{M} (k_1, a; k_2, b \rightarrow \psi) \mathcal{M}^*(k_1^\prime, c; k_2^\prime, d \rightarrow \psi) .
\end{equation}
To simplify this equation, note firstly that at leading order in $N$,
we may restrict the summation above to two particle states. Next, we
use the fact that the theory is invariant under the combined time
reversal times parity discrete symmetry, so that $\mathcal{M}(a
\rightarrow b)^* = \mathcal{M}(b \rightarrow a)$. Thus, the
requirement of unitarity becomes
\begin{multline}
\label{eq:unitary}
2 \mathrm{Im} \mathcal{M}(k_1, a; k_2, b \rightarrow k_1^\prime, c; k_2^\prime, d) = 
 \sum_{e,f} \iota_{ef}\int \frac{d^3q_1}{(2 \pi)^3 2 E_1} \frac {d^3 q_2}{(2 \pi)^3 2 E_2} (2 \pi)^4 \delta^4 (q_1 + q_2 - p_1 - p_2) \\
\times \mathcal{M} (k_1, a; k_2, b \rightarrow q_1, e; q_2, f) \mathcal{M}^*(k_1^\prime, c; k_2^\prime, d \rightarrow q_1, e; q_2, f) .
\end{multline}
where the identical particles factor $\iota_{ef}$ is equal to $1/2$ if
$e=f$ and is unity otherwise. It is now trivial to check
unitarity. First, notice that we can express the one-loop correction
$\Sigma(s)$ as
\begin{equation}
\label{eq:normalSigma}
\Sigma(p^2) 
= - {1 \over 32 \pi^2} \left[ \int_0^1 dx \log | m^2 - p^2 x (1-x) | - i \pi \sqrt{1 - \frac{4 m^2}{p^2}}\theta(p^2 - 4 m^2) \right].
\end{equation}
Therefore, since both $t$ and $u$ are negative, the imaginary part on the left hand side of Eq.~\eqref{eq:unitary} is given by
\begin{equation}
2 \mathrm{Im} \mathcal{M}(k_1 a; k_2 b\rightarrow k_1^\prime c; k_2^\prime d) = 
\frac{\lambda^2}{16 \pi N} 
\frac{ \sqrt{1 - \frac{4 m^2}{s}} \delta_{ab} \delta_{cd} }
{\left| 1 + \lambda \Sigma(s)  \right|^2 }.
\end{equation}
On the other hand, the sum over $e, f$ on the right hand side of
Eq.~\eqref{eq:unitary} is enhanced by one power of $N$ when the
scattering is in the $s$ channel. Thus, to leading order, the sum is
given by
\begin{multline}
\sum_{e,f} \iota_{ef}\int \frac{d^3q_1}{(2 \pi)^3 2 E_1} \frac {d^3 q_2}{(2 \pi)^3 2 E_2} (2 \pi)^4 \delta^4 (q_1 + q_2 - p_1 - p_2) 
\mathcal{M} (k_1, a; k_2, b \rightarrow q_1, e; q_2, f) \mathcal{M}^*(k_1^\prime, c; k_2^\prime, d \rightarrow q_1, e; q_2, f) \\
= \frac{N}{2} \delta_{ab} \delta_{cd} 
\int \frac{d^3 k_1}{(2 \pi)^3 2 E_1} \frac{d^3 k_2}{(2 \pi)^3 2 E_2} (2 \pi)^4 \delta^4(k_1 + k_2 - p) \left| {\lambda \over N}{1 \over  1 + \lambda \Sigma(s)} \right|^2 
= \frac{\lambda^2}{16 \pi N} \delta_{ab} \delta_{cd} \sqrt{1 - \frac{4 m^2}{s}} \left| {1 \over  1 + \lambda \Sigma(s)} \right|^2 ,
\end{multline}
where $s = p^2$.  Consequently, we see that the $S$-matrix of the
theory is unitary to leading order in $N$ on the two particle subspace
of the Hilbert space. Notice that this argument was sensitive only to
the imaginary part of $\Sigma(p^2)$. We will see that in the Lee-Wick
case, the real part of the one-loop correction is changed but the
imaginary part (for two particle final states) is the same. Since the
only nontrivial imaginary part is associated with the $\sigma$
propagator it should be clear that unitarity also holds for the higher
particle parts of the Hilbert space .
\end{widetext}

\subsection{Time dependence of Two Particle Scattering Processes}

In preparation for our discussion of acausal processes in the Lee-Wick
$O(N)$ model, we review some aspects of the time
dependence of the two particle scattering amplitude in the normal $O(N)$
model. To simplify the discussion we will assume $\lambda \ll 1$ and work to one-loop order in
perturbation theory. This approximation allows us to do explicit
computations while retaining the salient
features of the causal structure of the time dependent amplitude. Using the results of section
\ref{sec:kinematics},  the transition amplitude $
\langle \psi_\mathrm{out} | \psi_\mathrm{in} \rangle$ is given by
Eq.~\eqref{keyresult}:
\begin{equation}
\langle\psi_{{\rm out}}|\psi_{{\rm in}}\rangle
=\int \frac{d^4q}{(2\pi)^4} e^{-iq w}\hat
F(q)\hat G(q) \Gamma_s^{(4)}(q^2),
\end{equation}
where $w=z_0'-z_0$. The four point function $\Gamma_s^{(4)}(q^2)$ can
be deduced from Eq.~\eqref{eq:scatt} by expanding in the small
parameter $\lambda$. 
We ignore the tree-level amplitude as it
leads to trivial time dependence of the amplitude,
  $\exp(-i(p_1+p_2)\cdot w)$ times a function localized about $w=0$. The one-loop four
point function describing $(a, a) \rightarrow (b, b)$ scattering is
given by
\begin{equation}
\Gamma_s^{(4)}(q^2) = {-i \lambda^2 \over 32 \pi^2 N} \int_0^1 dx \log \left({m^2 - q^2 x(1-x) - i \epsilon \over \mu^2} \right).
\end{equation}
Thus, the transition amplitude is
\begin{widetext}
\begin{equation}
\langle\psi_{{\rm out}}|\psi_{{\rm in}}\rangle = { - i \lambda^2 \over  32 \pi^2 N} \int_0^1 dx \int {d^4 q \over (2 \pi)^4 } \hat F(q) \hat G(q) e^{i \vec q \cdot \vec w}\left( {1 \over -i w^0} {d \over dq^0} \right) \left( e^{-i q^0 w^0} \right) \log (m^2 -q^2 x(1-x) - i \epsilon),
\end{equation}
where we have introduced a derivative with respect to
$q^0$. Integrating by parts, this derivative acts on $\hat F^(q)$,
$\hat G(q)$ and the logarithm. Since the functions $\hat F(q)$ and
$\hat G(q)$ are slowly varying, we will only keep the term where the
derivative acts on the logarithm. Therefore the amplitude can be
written as
\begin{equation}
\langle\psi_{{\rm out}}|\psi_{{\rm in}}\rangle \simeq { - \lambda^2 \over  32 \pi^2 N w^0} \int_0^1 dx \int {d^4 q \over (2 \pi)^4} \hat F(q) \hat G(q) e^{-i q \cdot w} {2 q^0  \over  q^2 - m^2(x) + i \epsilon} , 
\end{equation}
where $m(x) = m / \sqrt{x(1-x)}$.  Introducing an integration over a variable $s$ to write the propagator as a phase  gives,
\begin{equation}
\langle\psi_{{\rm out}}|\psi_{{\rm in}}\rangle \simeq {  \lambda^2 \over 32 \pi^2 Nw^0} \int_0^1 dx \int _0^{\infty} ds \int {d^4 q \over (2 \pi)^4 } \hat F(q) \hat G(q) e^{-i  q \cdot w} e^{is(q^2-m^2(x)+i\epsilon)}2q^0.
\end{equation}

As before, we use the stationary phase approximation to evaluate
the various integrations. The stationary point for the integrations
over the components of $q$ is located at $q=w/(2s)$ and performing
these integrations gives (up to a constant phase)
 
 \begin{equation}
\langle\psi_{{\rm out}}|\psi_{{\rm in}}\rangle \simeq {\lambda^2 \over 16 \pi^2 N} {1\over (2\pi)^2}\int_0^1 dx\int_0^{\infty}ds\left( {1 \over 2 s}\right)^3 e^{-i(w^2/(4s)+m^2(x)s-i\epsilon s)} \hat F(w/(2s)) \hat G(w/(2s) ) .
\end{equation} 
Next the $s$ integration is performed using stationary phase.  The stationary point is at $s=\sqrt{w^2}/(2m(x))$ and we find that
\begin{equation}
\langle\psi_{{\rm out}}|\psi_{{\rm in}}\rangle  \simeq {\lambda^2 \over 32 \pi^2 N } {1 \over (2\pi)^{3/2}}\left({1 \over \sqrt{w^2}}\right)^{5/2}\int_0^1 dx (m(x))^{3/2}{\hat F}(m(x) w/ \sqrt{w^2}){\hat G}(m(x) w/ \sqrt{w^2}) e^{-im(x)\sqrt{w^2}} .
\end{equation}
Finally, we have to do the $x$ integral. We use the method of
stationary phase once again. The stationary point is at $x = 1/2$ and
the transition amplitude is
\begin{equation}
\langle\psi_{{\rm out}}|\psi_{{\rm in}}\rangle = { \lambda^2 \over 64 \pi^3 N} {m \over (\sqrt{w^2})^3} e^{-2 i m \sqrt{w^2}} \hat F(2mw/\sqrt{w^2}) \hat G(2mw/\sqrt{w^2}).
\end{equation}
Recall the functions ${\hat F}(q)$ and ${\hat G}(q)$ have support
  at $q=p_1+p_2$ and $q=p_1'+p_2'$, respectively. The amplitude is
  appreciable only $p_1+p_2\approx  p_1'+p_2' \approx 2mw/\sqrt{w^2}$.
Since the  energy is positive, and choosing the CM frame, the above can be rewritten as
\begin{equation}
\label{eq:normallooptime}
\langle\psi_{{\rm out}}|\psi_{{\rm in}}\rangle =\theta(w^0) { \lambda^2 \over 64 \pi^3 N} {m \over (w^0)^3} e^{-2 i m w^0} \hat F(2mw/\sqrt{w^2}) \hat G(2mw/\sqrt{w^2}).
\end{equation}
It is worth comparing this expression with the transition amplitude in
the case where the scattering is mediated by a resonance,
Eq.~\eqref{eq:normtreetime}. In that case there an exponential decay
due to the width of the resonance, which is absent in
Eq.~\eqref{eq:normallooptime} because the mediators are stable. In the
tree-level case the scattering is mediated by one particle; its
wave packet spreads out like $(w^0)^{3\over2}$. In the loop case the
presence of two wave packets leads to a power-law fall off of the
amplitude as $(w^0)^3$. The $\theta$-function in
Eq.~\eqref{eq:normallooptime} indicates that the decay particles
appear at times after the collision.

\end{widetext}

\section{The Lee-Wick $O(N)$ Model}

Let us now move on to study the Lee-Wick $O(N)$ model. We begin by
discussing the Lagrangian of the model before moving on to examine the
loop structure. Once the loop structure is understood at leading order
in the $1/N$ expansion, we will compute the two particle scattering to
leading order in $1/N$ and to all order in $\lambda$. We will use
these results to demonstrate unitarity of the theory. Finally, we will
explicitly compute the time dependence of one-loop scattering
processes, demonstrating aspects of the acausality of the model.

\subsection{The Lagrangian}

The theory is the usual $O(N)$ model, augmented with a higher derivative term. The Lagrangian is given by
\begin{eqnarray}
&&{\cal L}={1 \over 2}\partial_{\mu} \hat \phi^a \partial^{\mu} \hat \phi^a - {1 \over 2 M^2} \partial^2 \hat \phi^a \partial^2 \hat \phi^a-{ 1 \over 2} m_0^2 \hat \phi^a \hat \phi^a
\nonumber \\
 &&-{\lambda \over 8 N} \left(\hat \phi^a \hat \phi^a \right)^2,
\end{eqnarray}
where, as in the normal $O(N)$ model, the fields $\hat \phi^a$ are
scalar fields in the fundamental representation of the group
$O(N)$. We can remove the higher derivative term from the Lagrangian
by introducing $N$ scalar fields $\tilde \phi^a$. Then an equivalent
Lagrange density is
\begin{multline}
{\cal L}={1 \over 2}\partial_{\mu} \hat \phi^a \partial^{\mu} \hat \phi^a -{ 1 \over 2} m_0^2 \hat \phi^a \hat \phi^a - \tilde \phi^a \partial^2 \hat \phi^a \\
+ {1 \over 2} M^2 \tilde \phi^a \tilde \phi^a -{\lambda \over 8 N} \left(\hat \phi^a \hat \phi^a \right)^2.
\end{multline}
We may diagonalize the derivative terms by defining $\phi^a = \hat \phi^a + \tilde \phi^a$ and performing an integration by parts. The Lagrangian becomes
\begin{multline}
{\cal L}={1 \over 2}\partial_{\mu} \phi^a \partial^{\mu} \phi^a -{ 1 \over 2} m_0^2  (\phi^a - \tilde \phi^a) (\phi^a -\tilde \phi^a)  - {1 \over 2} \partial_\mu \tilde \phi^a \partial^\mu \tilde \phi^a \\
+ {1 \over 2} M^2 \tilde \phi^a \tilde \phi^a -{\lambda \over 8 N} \left[ \left(\phi^a - \tilde \phi^a \right) \left( \phi^a - \tilde \phi^a \right) \right]^2.
\end{multline}
This Lagrangian has a simple interpretation. There are $N$ normal
scalar fields $\phi^a$ and $N$ Lee-Wick scalar fields $\tilde \phi^a$;
these fields have quartic interactions.  Note that the Lee-Wick
scalars $\tilde \phi^a$ can decay to three $\phi^a$ quanta provided
that $\tilde \phi^a$ is heavy enough. We will assume this decay
channel is open so that the width $\Gamma$ of the Lee-Wick scalars is
non-zero.  There is mass mixing between the normal and Lee-Wick
scalars which can be removed by a symplectic transformation on the
fields or treated as a perturbation when the ordinary scalars are very
light compared with the Lee-Wick scalars. Neglecting the mass of the
ordinary scalars, for large $N$ and small coupling $\lambda$,
\begin{equation}
\Gamma\simeq
 {\lambda^2 M \over  2^{10}\pi^3 N }. 
\end{equation}
To obtain a unitary $S$ matrix it is crucial that the  $\tilde
\phi$ propagator  have poles at complex energy, as in
\eqref{fullLWprop}. For this reason, even though the width is order $1/N$ it will be
important to retain it in intermediate steps in our calculations.

As was the case in the normal $O(N)$ model, it is convenient to
introduce a non-dynamical scalar field $\sigma$ so that the Lagrangian
may be written as
\begin{multline}
{\cal L}={1 \over 2}\partial_{\mu} \phi^a \partial^{\mu} \phi^a -{ 1 \over 2} m_0^2 \left(\phi^a - \tilde \phi^a \right) \left( \phi^a - \tilde \phi^a \right) + {N \over 2 \lambda} \sigma^2\\
 - {1 \over 2} \partial_\mu \tilde \phi^a \partial^\mu \tilde \phi^a 
+ {1 \over 2} M^2 \tilde \phi^a \tilde \phi^a  -{1 \over 2} \sigma \left(\phi^a - \tilde \phi^a \right) \left( \phi^a - \tilde \phi^a \right).
\end{multline}
In this formulation of the theory, it is straightforward to establish
a power-counting argument which shows that at leading order in $1/N$,
the only relevant graphs are the $\sigma$ tadpole and the 1PI $\sigma$
self-energy $\Sigma_0(q^2)$. We can absorb the effects of the $\sigma$
tadpole by replacing $m_0^2$ by $m^2$. In the following, we will treat
$m$ as a small parameter. It remains to compute the self-energy,
$\Sigma_0(q^2)$ of $\sigma$ to leading order in $1/N$.

\subsection{Loops}

Before we embark on the computation of the self-energy, let us pause
for a moment to consider the properties of the Lee-Wick resonances. We
are familiar with the properties of ordinary resonances in quantum
field theory. One familiar fact is the importance of resuming the
width of a resonance in order to avoid the appearance of spurious
poles in Feynman graphs. This resummation changes the analytic
structure of the theory in a manner consistent with the
non-perturbative information of the Lehmann representation. Similarly,
resummation of the width of Lee-Wick resonances changes the analytic
structure of the theory in crucial ways.

At tree level, the Lee-Wick propagator is
\begin{equation}
\tilde D(p^2) =  -{i \over p^2 - M^2}.
\end{equation}
At loop level, the particle develops a width. Just as in the example
we discussed earlier, in Sec.\ref{sec:LWresBeh}, the loop-corrected
Lee-Wick propagator is given by
\begin{equation}
\label{eq:LWpropON}
\tilde D(p^2) =  -{i \over p^2 - M_c^2} - {i \over p^2 - M_c^{*2}} 
+ {i \over \pi}
 \int_{9 m^2}^\infty ds
 {\rho(s) \over p^2 - s + i \epsilon},
\end{equation}
where $M_c^2=M^2+iM\Gamma$. We shall use this form of the propagator
to compute the $\sigma$ self-energy even though the corrections due to
the width are formally subdominant in the $1/N$ expansion. The
subdominant corrections modify the analytic structure and it is this
modification that allows the theory to be unitary.

The poles present in the Lee-Wick propagator are in unusual locations
in the complex $p^2$ plane, so we must take care to define the contour
of integration in Feynman graphs appropriately. We must understand how
to define expressions such as
\begin{equation}
\label{eq:exampleInt}
I = \int {d^4 p \over (2 \pi)^4} {-i \over (p+q)^2 - M_1^2} {-i \over p^2 - M_2^2},
\end{equation}
where $M_1$ and $M_2$ may be complex masses, either in the upper or lower half plane of the Feynman integration.
Let us consider the $p^0$ integral.  The integrand has four poles.  Two of the poles are located at
\begin{equation}
\label{pole1}
p^0=\pm \sqrt{ {\vec p}^2+M_2^2}.
\end{equation}
The location of the other two poles depends on the value of the
external four momentum $q$. 
For time-like $q$ we can go to a frame where ${\vec q}=0$ and these two poles are located at,
\begin{equation}
\label{pole2}
p^0=-q^0 \pm \sqrt{ {\vec p}^2+M_1^2}.
\end{equation}

The contour Lee and Wick suggested is such that, once the Green's
  functions are computed by Fourier transform from momentum space to space-time, there is no
exponential growth in time, and can be described as follows. Consider the
position of the poles as a function of the coupling $\lambda$ present
in the theory. At $\lambda = 0$ the widths vanish, so $M_1$ and $M_2$
are real masses. Then the contour is defined to be the usual Feynman
contour. As $\lambda$ increases away from zero the Lee-Wick
  particles become unstable, the poles on the real line become 
  complex pairs of poles that move away from the real axis. The
  Lee-Wick prescription is to deform the contour, as $\lambda$
  increases from zero,  so that the complex
poles do not cross the contour; a pole which was initially below the
contour remains below the contour, for example. 
\begin{figure}[htbp] 
    \centering
    \includegraphics[width=3in]{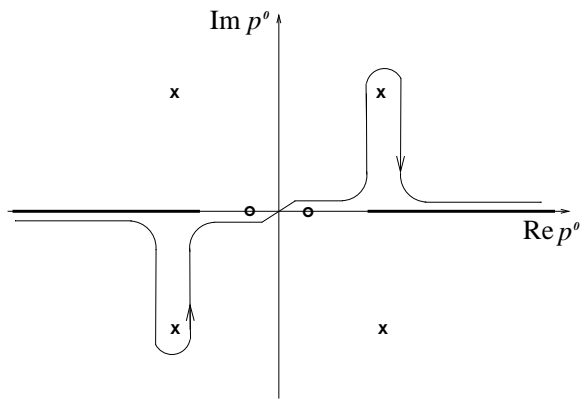}
    \caption{Contour given by the Lee-Wick prescription for  
integration in
      the  complex $p^0$ plane. The crosses
      denote the poles at  $p^0=\pm\sqrt{\vec p^2+M_c^2}$ and at
      $p^0=\pm\sqrt{\vec p^2+M_c^{*2}}$
   and the circles those at $p^0=-q^0\pm\sqrt{(\vec p+\vec
     q)^2+m^2}$. The heavy line denotes the cuts on the real axis
   starting at $\pm3m$. The contour of integration is  deformed as the
   interactions are turned on and the LW poles move into the complex
   plane so that the complex
poles do not cross the contour.}
    \label{fig:k0-plane-6}
\end{figure}
If the external momentum is in the unphysical region this
  prescription is unambiguous. For example, for the integral in \eqref{eq:exampleInt}, one can start with
  $|q^0|<|M_1+M_2|$. As the momentum is varied (for fixed,
  non-zero $\lambda$) poles may cross a contour. However, the
  integral can still be defined by deforming the contour so as to
  avoid the pole. This
leads to a well-defined contour unless poles pinch it. The pinching
occurs when a pole in 
Eq.~\eqref{pole1} coincides with one in Eq.~\eqref{pole2}, and signals
the presence of a singularity, usually a branch cut, in the integral
$I$ (as a function of $q$). An additional prescription is required to
define the integral in this case. 

We need a prescription only when the new singularity occurs for real
valued energy $q^0$. This may occur if one propagator carries mass
$M_1^2$ while the other one has mass $M_1^{*2}$ --- that is, we could have
$M_2^2 = M_1^{*2}$ in Eq.~\eqref{eq:exampleInt}.
Then it is easy to see that when $q^0$ satisfies the equation
\begin{equation}
\left(q^0\right)^2=2({\vec p}^2+{\rm Re M_1}^2 ) +2|{\vec p}^2+M_1^2| 
\end{equation}
two of the poles in Eq.~(\ref{pole1}) and Eq.~(\ref{pole2}) overlap
and the contour is pinched.  The CLOP prescription is as follows:
Define the Feynman integral by taking the masses $M_1^2$ and $
M_1^{*2}$ to be unrelated complex mass parameters so that the poles do
not overlap. At the end of the calculation  impose the condition
that $ M_2^{2}$ is the complex conjugate of $M_1^2$. With this
prescription, the self-energy $\Sigma_0$ is unambiguously defined
  and Lorentz invariant, and may be
computed using standard methods. In particular, the contour we have
chosen allows us to Wick-rotate the integral in
Eq.~\eqref{eq:exampleInt} and in all Feynman integrals we will
encounter. In the following, we will compute the integrals in
dimensional regularization and discard the divergent pieces that are
proportional to $1/(d-4)$ in addition to the finite pieces involving
the logarithm of $4 \pi$ and Eulers constant. In the Lee-Wick O(N)
model they cancel, since the theory is finite by naive power counting
in the higher derivative formulation.\footnote{Except for one loop
    self-energy graphs, which give a trivial log divergence mass shift.}

\subsection{Computation of the Self-Energy}

We now embark on the explicit computation of the self-energy. We
define $\Sigma = N \Sigma_0$ as before. There are various graphs
contributing. The graph involving only the normal particles reproduces
the self-energy in the normal $O(N)$ model,
Eq.~\eqref{eq:normalSigma}. To simplify the notation, we define a
function
\begin{multline}
F(M_1^2, M_2^2, q^2) =  {i \over 2}\int {d^d p \over (2 \pi)^d} {i \over (p+q)^2 - M_1^2} \\
\times {i \over p^2 - M_2^2} .
\end{multline}
Discarding the divergent piece that is proportional to $1/(d-4)$ and
the finite constant pieces involving the logarithm of $4\pi$ and
Euler's constant we find
\begin{multline}
F(M_1^2, M_2^2, q^2) =-{1 \over 32 \pi^2} \\
\times \int_0^1 dx \log \left({x M_1^2 + (1-x) M_2^2 -x (1-x) q^2 }\right).
\end{multline}
It is easy to verify that the only candidate singularity of  $F(M_1^2, M_2^2, q^2)$, as a function of the complex variable $q^2$,
is a branch cut with branch point at $q^2=(M_1+M_2)^2$. 
The $\sigma$ self-energy can then be expressed in terms of sums of
these functions evaluated at various arguments. Thus, the
contribution, $\Sigma_1$ of the normal particles to the total
self-energy, $\Sigma$, is
\begin{equation}
 \Sigma_1(q^2) = F(m^2, m^2, q^2).
\end{equation}
Since one of our main goals is to understand unitarity of the theory, we will focus on understanding any possible imaginary parts of $\Sigma$.

Next we consider graphs with both propagators being of the
  Lee-Wick field, \eqref{eq:LWpropON}. It is convenient to consider the contributions of the pole parts of
graphs involving the Lee-Wick particles separately from contributions
involving the spectral density $\rho(s)$. Firstly, consider the terms
involving only the Lee-Wick poles. Following the CLOP prescription
  we use different complex masses for the two propagators, with
  $M_1^2=M_c^2+i\delta$ and $M_2^2=M_c^2$. The loop integral is
\begin{widetext}
 \begin{align}
 \Sigma_2(q^2) &= -{i \over 2}\int  \frac{d^d p}{(2 \pi)^d} \left[
  \frac{1}{p^2 - M_c^2 - i \delta}
+  \frac{1}{p^2 - M_c^{*2} + i \delta}\right]\;\;\left[
\frac{1}{(p+q)^2 - M_c^2} +
\frac{1}{(p+q)^2 - M_c^{*2}}\right]\nonumber\\
&= F(M_c^2+ i\delta, M_c^2,q^2) + F( M_c^{*2}-i\delta,  M_c^{*2},q^2) + F(M_c^2 + i \delta,
 M_c^{*2},q^2) + F(M_c^{*2} - i \delta, M_c^2,q^2)~.
\end{align} 
It is easy to see that  $\Sigma_2$ is continuous
across the real line. The CLOP prescription has effectively moved the
two branch points that would have occurred at $q^2=(M_c+M_c^*)^2$ away
from the real axis, by an amount of order $\delta$,
to $\sqrt{M_c^2+i\delta}+M_c^*$ and
$\sqrt{M_c^{*2}-i\delta}+M_c$. The remaining two terms appearing
in the self-energy have complex branch points even for $\delta=0$. Hence the
discontinuity across the real line vanishes, and this persists in the
limit that $\delta$ goes to zero. An explicit example may clarify
  this. The expression
$F(M_c^2 + i \delta, M_c^{*2}) + F(M_c^{*2} - i \delta, M_c^2)$ 
 contains the dangerous terms appearing in the Feynman integral in
which poles on opposite sides of the contour may pinch when $\delta =
0$ (and $q^2$ is real.) But this expression is explicitly real on the
real axis and analytic in a band of width $\sim\delta$ containing the
whole real axis:
\begin{multline}
F(M_c^2 + i \delta, M_c^{*2},q^2) + F(M_c^{*2}- i \delta, M_c^2 ,q^2) 
\\
= -{1 \over 32 \pi^2} \int_0^1 dx \left[ \log (x (M_c^2 + i \delta) + (1-x) M_c^{*2} -x (1-x) q^2) + \log (x (M_c^{*2}+ i \delta) + (1-x) M_c^2  -x (1-x) q^2) \right] \\
= -{1 \over 32 \pi^2} \int_0^1 dx \log \left[ \left|x (M_c^2 + i \delta) + (1-x) M_c^{*2} -x (1-x) q^2\right|^2  \right] .
\end{multline}
Since for real valued $q^2$ the imaginary part of this vanishes (there is no need to define
 this as a discontinuity) independent of $\delta\ne0$. In the limit as
 $\delta \rightarrow 0$ the imaginary part remains zero.  
In the remainder of this section, we will omit the parameter $\delta$
to simplify the equations.

In the next section we will use the self-energy to verify the
  unitarity of the $S$ matrix in this theory. It will be useful to
  write the result for the self-energy concisely. While the width is
  of order $1/N$, its presence is crucial in demonstrating that $\Sigma_2$
  is real. But once we established  that the CLOP defined  $\Sigma_2$ is
  real we can neglect the width and give a simple expression for the
   self-energy:
\begin{equation}
\Sigma_2(q^2) = - {1 \over 16 \pi^2} \int_0^1 dx \log |M^2 - x (1-x) q^2 |^2.
\end{equation}

Next, we compute the Feynman integrals involving the Lee-Wick pole and the normal particle.
We find that the self-energy in this case is given by
\begin{align}
\label{eq:sigma3}
\Sigma_3(q^2) &= {i \over 2} \int  \frac{d^d p}{(2 \pi)^d} \left[ \frac{1}{p^2 -m^2} \left(
\frac{1}{(p+q)^2 - M_c^2} + \frac{1}{(p+q)^2 - M_c^{*2}} \right) 
+ \frac{1}{(p+q)^2 -m^2} \left( \frac{1}{p^2 - M_c^2}
+ \frac{1}{p^2 -  M_c^{*2}} \right) \right]
\nonumber \\
&=-2F(m^2,M_c^2,q^2)-2F(m^2,M_c^{*2},q^2)~.
\end{align}
The branch points are both off the real axis, at $q^2=(M_c+m)^2$ and
$(M^*_c+m)^2$. As above, since $\Sigma_3$ is real we can neglect the
width and write, concisely:
\begin{equation}
\Sigma_3(q^2)={1 \over 16 \pi^2} \int_0^1 dx \log |x M^2+(1-x)m^2 - x (1-x) q^2|^2.
\end{equation}

Now we turn to terms involving the spectral density $\rho(s)$. Terms in the self-energy
involving products of the Lee-Wick poles and the spectral density give
\begin{align}
\Sigma_4 &=\frac{i}{\pi}\int_{9m^2}^{\infty} ds \rho(s) \int \frac{d^d p}{(2 \pi)^d} \frac{1}{p^2 - s + i \epsilon}
\left( \frac{1}{(p+q)^2 -  M_c^{*2} }+ \frac{1}{(p+q)^2 - M_c^2}
\right)\nonumber\\
&=-\frac{2}{\pi} \int_{9m^2}^{\infty} ds \rho(s)\left[F(s,M_c^2,q^2)+F(s,M_c^{*2},q^2)\right]~.
\end{align}
The sum of $F$ functions is similar to that appearing in 
Eq.~\eqref{eq:sigma3}: evidently this is also  real and so the integral
against $\rho$ is real. In the narrow resonance approximation, we find that self-energy due to these terms is
\begin{equation}
\Sigma_4 = {1 \over 16 \pi^2} \int_0^1 dx \log |M^2 - x(1-x) q^2 |^2 .
\end{equation}

Finally, there are terms involving the spectral density $\rho$ and the
normal pole, and involving a double integral over two powers of
$\rho$.  These terms do lead to an imaginary part, describing real
scattering from two particle states into four or six particle
states. It is important to understand that the imaginary parts arising
from these expressions involve final states containing only normal
particles. In the narrow resonance approximation, we find that these
terms lead to a contribution to the self-energy given by
\begin{multline}
\Sigma_5 = -{1 \over 32 \pi^2} \int_0^1 dx \log |M^2 - x (1-x) q^2| + {i \pi \over 32\pi^2} \sqrt{1 - {4 M^2 \over q^2}} \theta (q^2 - 4 M^2) \\
- {1 \over 16 \pi^2} \int_0^1 dx \log | x M^2+(1-x)m^2 - x (1-x) q^2 | + { i \pi \over 16 \pi^2} \left( 1 - {(m+M)^2 \over q^2 }\right) \theta (q^2 - (m+M)^2) .
\end{multline}

In total, we find an explicit expression for the self-energy which is simple when we treat the width $\Gamma$ to be negligible compared to the mass $M$:
\begin{multline}
\label{eq:LWselfenergy}
\Sigma(q^2) = -{1 \over 32 \pi^2} \left[
\int_0^1 dx 
\log  { | x(1-x) q^2 | | M^2 - x (1-x) q^2 | \over |x M^2 - x (1-x) q^2 |^2 }
- i \pi \theta(q^2) 
\right. \\ \left.
- i \pi \sqrt{1 - {4 M^2 \over q^2 }} \theta(q^2 - 4M^2) 
- 2 i \pi \left( 1 - {M^2 \over q^2} \right) \theta (q^2 - M^2)\right],
\end{multline}
where we have neglected the normal mass $m$.
Notice that the width $\Gamma$ of the Lee Wick resonances does not appear in this result.  It was important for defining the contour for the loop integration but  not in the final form of the answer. It will also play a role in our understanding of the unitarity of the $S$ matrix.
\subsection{Unitarity}

Unitarity of the $S$ matrix is equivalent to requiring $ i(T^{\dagger}-T)=T^{\dagger}T$. We consider two particle matrix elements of the right and left hand side of this equation and verify their equality to leading order in $1/N$. For convenience, we restate the
requirement of unitarity for the amplitude describing scattering of a two particle
state into a two particle state:
\begin{equation}
\label{LWunitarity}
i \left(\mathcal{M}(k_1^\prime, c; k_2^\prime, d \rightarrow k_1, a; k_2, b)^* - \mathcal{M}(k_1, a; k_2, b \rightarrow k_1^\prime, c; k_2^\prime, d) \right) = \sum_\psi \mathcal{M} (k_1, a; k_2, b \rightarrow \psi) \mathcal{M}^*(k_1^\prime, c; k_2^\prime, d \rightarrow \psi), 
\end{equation}
where $\mathcal{M}(k_1, a; k_2, b \rightarrow k_1^\prime, c;
k_2^\prime, d)$ is the amplitude for the two particle
scattering. Previously, in our discussion of unitarity of scattering
in the normal $O(N)$ model, we argued that the only allowed final
state at leading order is the two particle final state. However, the
situation is different in the Lee-Wick $O(N)$ model. At leading order,
two, four and six particle final states are accessible. Intuitively,
this is because we can create Lee-Wick resonances which subsequently
decay into three normal particles with unit probability. This is
  the reason it was necessary to retain the width of the Lee-Wick
  resonances, even if it is subleading in $1/N$; after all, for a
  non-zero width, however small, given enough time the unstable
  ``particle'' will decay. Therefore we
will have to include these additional final states in the sum of the
right hand side of Eq.~\eqref{LWunitarity}. In Eq.~(\ref{LWunitarity})
the initial and final states and the intermediate states $\psi$ only
involve the stable ordinary particles.  Even though the propagator for
the Lee-Wick resonances contains poles at $p^2=M_c^2$ and
$p^2=M_c^{*2}$ these particles are not considered to be in the
spectrum of the theory.

The two particle scattering amplitude in the Lee-Wick theory is given
in terms of the self-energy by the same expression as in the normal
$O(N)$ model:
\begin{equation}
\label{LWamp}
{\cal M}(k_1, a; k_2, b \rightarrow k_1^\prime, c; k_2^\prime, d)=-{\lambda \over N}\left(  {\delta_{ab}\delta_{cd} \over 1+\lambda \Sigma(s)}+ \ldots \right),
\end{equation}
where $s = (k_1 + k_2)^2$ as usual, and the dots indicate the $t$ and
$u$ channel terms in addition to higher order terms in $1/N$. Since $t$
and $u$ are negative quantities for physical scattering, we find that
the left hand side of the unitarity relation Eq.~\eqref{LWunitarity} is
\begin{multline}
\label{LWunitLeft}
i \left(\mathcal{M}(k_1^\prime, c; k_2^\prime, d \rightarrow k_1, a; k_2, b)^* - \mathcal{M}(k_1, a; k_2, b \rightarrow k_1^\prime, c; k_2^\prime, d) \right) 
\\  
= {\lambda^2 \over 16 \pi N} {\delta_{ab}\delta_{cd} \over |1+\lambda \Sigma(s)|^2} \left( 
1+\sqrt{1 - {4 M^2 \over s }} \theta(s - 4 M^2) 
+ 2 \left( 1 - {M^2 \over s} \right) \theta (s - M^2) \right) .
\end{multline}
For simplicity, and without loss of generality, here and below we neglect the normal mass $m$.  Now we must
compute the right hand side of the unitarity relation. This is
straightforward when the state $| \psi \rangle$ in
Eq.~\eqref{LWunitarity} is a two particle state; in that case the sum
becomes an integral over the two body phase space of the normal
particles and the amplitude is simply the two-two scattering amplitude
of Eq.~\eqref{LWamp}. Since we are neglecting the mass of $\phi^a$ the
sum becomes
\begin{equation}
\label{LWunitR1}
\sum_{|\psi\rangle = | q_1, e; q_2, f \rangle } \mathcal{M} (k_1, a; k_2, b \rightarrow \psi) \mathcal{M}^*(k_1^\prime, c; k_2^\prime, d \rightarrow \psi)
=
\frac{\lambda^2}{16 \pi N} \delta_{ab} \delta_{cd} \left| {1 \over  1 + \lambda \Sigma(s)} \right|^2.
\end{equation}
We see that this part of the complete sum over accessible
final states $|\psi \rangle$ reproduces the first term in parentheses on the right hand side of
Eq.~\eqref{LWunitLeft}. The other two terms arise from four and six
particle final states. Consider first the process with a four particle
final state. To leading order in $N$, the two-four particle scattering
must contain an intermediate $\tilde \phi$ which then decays; the
Feynman graph is shown in Fig.~\ref{fig:two-to-four}. 
\begin{figure}[htbp] 
    \centering
    \includegraphics[width=3in]{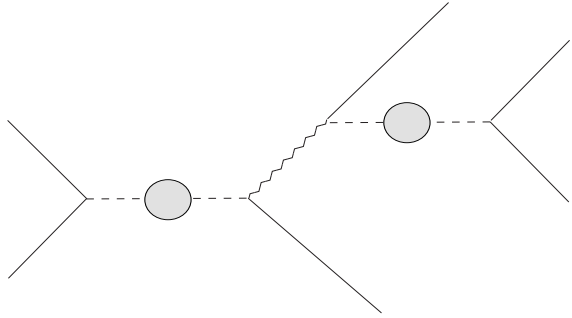}
    \caption{Feynman diagram for the two to four scattering amplitude
      proceeding through the decay of a Lee-Wick particle. The solid  
line denotes a ``normal''
particle, the zig-zag line a ``Lee-Wick'' particle, and the dashed
line with the shaded blob denotes the dressed $\sigma$ auxiliary field propagator.}
    \label{fig:two-to-four}
\end{figure}

Since we are summing over all
final states in the decay of the Lee-Wick resonance, the computation
reduces to computing the amplitude to make the intermediate state
containing a Lee-Wick  resonance and one normal particle, and integrating over
two particle phase space. This is familiar from the theory of ordinary
resonances. In more detail,
the four particle phase space integral can be organized into an integral
over the three body phase space for the Lee-Wick decay  and an
integral over the two particle phase space of the Lee-Wick and fourth
normal particle times an energy-momentum conserving delta function. The
result of the three particle phase space integral is the width of the
Lee-Wick particle; however, in the region of the two body phase space
where the Lee-Wick is nearly on-shell a factor of $1/\Gamma$ appears
from the Lee-Wick propagator. Thus, the whole computation reduces to
a simple integral over the two body phase space. The remaining parts
of the integrand can then be interpreted as the amplitude to create
the intermediate Lee-Wick particle plus one of the final state normal
particles.

The amplitude to create the intermediate state is given up to sign
by Eq.~\eqref{LWamp}. Thus, we can explicitly perform the sum over
four body phase space to find
\begin{equation}
\label{LWunitR2}
\sum_\mathrm{4 particle} \mathcal{M} (k_1, a; k_2, b \rightarrow \psi) \mathcal{M}^*(k_1^\prime, c; k_2^\prime, d \rightarrow \psi)
=
\frac{2\lambda^2}{16 \pi N} \delta_{ab} \delta_{cd} \left( 1 - {M^2 \over s} \right) \theta(s - M^2) \left| {1 \over  1 + \lambda \Sigma(s)} \right|^2.
\end{equation}
Similarly, the six particle phase space integral becomes an integral over the
two particle phase space of two intermediate Lee-Wicks. Near the region where
the Lee-Wicks are on-shell, there is an enhancement by $1 / \Gamma^2$. The result
is
\begin{equation}
\label{LWunitR3}
\sum_\mathrm{6 particle} \mathcal{M} (k_1, a; k_2, b \rightarrow \psi) \mathcal{M}^*(k_1^\prime, c; k_2^\prime, d \rightarrow \psi)
=
\frac{\lambda^2}{16 \pi N} \delta_{ab} \delta_{cd} \sqrt{ 1 - {4M^2 \over s}} \theta(s -  4M^2) \left| {1 \over  1 + \lambda \Sigma(s)} \right|^2.
\end{equation}
Since the sum of Eqs.~\eqref{LWunitR1}, \eqref{LWunitR2}
and~\eqref{LWunitR3} equals Eq.~\eqref{LWunitLeft} we have verified,
for two particle matrix elements, that $
i(T^{\dagger}-T)=T^{\dagger}T$.  

It is easy to extend this argument to show unitarity for any
  matrix element. To leading order in $1/N$ any  amplitude is given by
  a sum of skeleton diagrams with the propagators, including the
  $\sigma$ propagator, replaced by the full propagators. In the
  absence of a Kallen-Lehman decomposition, we cannot proceed with the
  usual cutting rules to show unitarity. For example, it is not
  obvious, if at all, how to set up a ``largest time
  equation~\cite{'t Hooft}.  However, one can still
  analyze individual graphs by cutting the diagrams. A cut through a
  $\sigma$  propagator is handled using the results for the $2\to2$
  amplitude demonstrated above. Cuts through normal particle
  propagators never produce an imaginary part: they are just as in the
  standard analysis and, since we only have skeleton graphs, these
  propagators are never on-shell. Finally there are ``cuts'' through
  the Lee-Wick propagators. These just correspond to taking the
  imaginary part of $\tilde D(p^2)$ in \eqref{eq:LWpropON}. The
  imaginary part of the sum of complex poles vanishes. We are left
  with the imaginary part of the integral over the spectral function
  $\rho(s)$. This has precisely the structure that a normal
  resonance in the standard unitarity analysis has, so it leads to the
  correct unitarity relation. In particular, it corresponds to a sum,
  in $T^\dagger T$, over intermediate three normal particle states.

Since the $S$ matrix provides a one-to-one map from the past to the
future in scattering experiments, the existence of a well-defined $S$
matrix is enough to show that there can be no paradoxes in these
scattering processes. Nevertheless, the theory is acausal as we shall
now explore.

\end{widetext}

\subsection{Time Dependence: Acausality}

To study the time dependence of scattering in the Lee-Wick theory, we
will work to one-loop order in perturbation theory. The graph
containing normal particles reproduces the transition amplitude of the
normal $O(N)$ model, shown in Eq.~\eqref{eq:normallooptime}. Our main
focus is on the acausal behavior associated with poles in the upper
half plane.  All the acausality decays exponentially with time except
for the case where in the loop one of the poles is at $M_c^2$ and the
other at $M_c^{*2}$. Then for real incoming momentum one can create an
on shell configuration with two Lee-Wick resonances and this leads to
acausal behavior that falls off with a power of time. In this section
we calculate this power law acausal behavior.  The part of the four
point function with Lee-Wick poles at $M_c^2$ and $M_c^{*2}$ is
\begin{equation}
\Gamma_s^{(4)}(q^2) =  {-i \lambda^2 \over 16 \pi^2 N} \int_0^1 dx \log ( M^2 - i (1-2x) M \Gamma - x(1-x) q^2 ).
\end{equation}
Since the sign of the $M \Gamma$ term in the logarithm changes sign
over the region of integration, it is convenient to break the integral
into two terms as
\begin{align}
&\Gamma_s^{(4)}(q^2) =  {-i \lambda^2 \over 16 \pi^2 N} \int_0^{1 \over 2} dx \left[ \log ( M^2 - i (1-2x) M \Gamma    \right. \nonumber  \\
&- x(1-x) q^2 )+ \log ( M^2 + i (1-2x) M \Gamma \left. - x (1-x) q^2 ) \right]. 
\end{align}
We make the same assumptions as before.  In particular the functions
${\hat F}(q)$ and $\hat G (q)$ are taken to be slowly varying and to
have support around $q^0= 2M$ and ${\vec q}=0$. We find it convenient
to decompose the transition amplitude as
\begin{equation}
\langle \psi_\mathrm{out} | \psi_\mathrm{in} \rangle =\langle \psi_\mathrm{out} | \psi_\mathrm{in} \rangle_+ +\langle \psi_\mathrm{out} | \psi_\mathrm{in} \rangle _-
\end{equation}
\begin{widetext}
\noindent
where
\begin{equation}
\langle \psi_\mathrm{out} | \psi_\mathrm{in} \rangle _{\pm} \simeq -{\lambda^2 \over 8 \pi^2 N w^0} \int_0^{1 \over 2}dx \int {d^4 q \over (2 \pi)^4} \hat F(q) \hat G(q) e^{-i q \cdot w} \left[ {q^0 \over q^2 -M(x)^2 \pm i {1 - 2x \over x(1-x) } M \Gamma} \right].
\end{equation}
Here, $M(x)=M/\sqrt{x(1-x)}$. We now put the denominator of the
propagator into an exponential by introducing an integration over a
proper time variable $s$,
\begin{equation}
\langle \psi_\mathrm{out} | \psi_\mathrm{in} \rangle_{\pm}\simeq {\pm i \lambda^2 \over 8 \pi^2 N w^0} \int_0^{1\over 2}dx \int_0^{\infty} ds \int {d^4 q \over (2\pi)^4} \hat F( q) \hat G( q) q^0 \exp \left(-i  q \cdot  w \pm is(q^2-M(x)^2)-s(1-2x)M\Gamma/(x(1-x))\right).
\end{equation}
It is now straightforward to successively do the $q$, $s$ and $x$
integrations using the stationary phase approximation. The stationary
points are at $ q=\pm w/(2s)$, $s=\sqrt{w^2}/(2M(x))$ and $x=1/2$ and
we find that (up to an overall constant phase)
\begin{equation}
\langle \psi_\mathrm{out} | \psi_\mathrm{in} \rangle_{\pm} \simeq {\lambda^2M \over 32 \sqrt{ w^2 }^3\pi^3 N } e^{\mp i 2 M \sqrt{w^2}}{\hat F}(\pm2Mw/\sqrt{w^2}){\hat G}(\pm2Mw/\sqrt{w^2}).
\end{equation}
Given where the functions $\hat F$  and $\hat G$ have support we can rewrite this as,
\begin{equation}
\label{eq:powerlawtime}
\langle \psi_\mathrm{out} | \psi_\mathrm{in} \rangle_{\pm} \simeq {\theta (\pm w^0)\lambda^2M \over 32| w^0| ^3\pi^3 N } e^{- i 2 M w^0}{\hat F}(\pm2Mw/\sqrt{w^2}){\hat G}(\pm2Mw/\sqrt{w^2}).
\end{equation}
We have checked by explicit calculation that the other one loop
contributions are exponentially suppressed in $|w^0|$ and so for very
large $ |w^0|$ the power law term that falls off as $1/|w^0|^3$,
displayed above, dominates the acausality in the one-loop contribution
to $\langle \psi_\mathrm{out} | \psi_\mathrm{in} \rangle$. Note that
Eq.~(\ref{eq:powerlawtime}) has a very different behavior than one
would expect based on the example of single Lee-Wick resonant exchange
that we discussed earlier. It is not exponentially suppressed for
large times and contains both acausal and causal pieces.

\end{widetext}


\section{Concluding Remarks}

We have studied the Lee-Wick $O(N)$ model and argued that the
prescription of Lee and Wick and Cutkowski {\it et. al.} yields an
S-matrix for this theory that is unitary and Lorentz invariant in
large $N$.  This suggests that, even though the theory is not causal,
there will not be paradoxical behavior in scattering experiments.

In this model we demonstrated, by explicit calculation, some of the
acausal behavior in two-two scattering of the ordinary scalars that
arises from virtual ``Lee-Wick particles.'' The Lee-Wick $O(N)$ model
presents a playground to examine the consistency of theories where
causality emerges only for long enough times and low enough
energies. There are other theories that are worth exploring for this
purpose. For example, there are two dimensional models that can be
solved exactly and it would be interesting to see if Lee-Wick versions
of some of these theories are also soluble and, if so, explore their
properties.




\begin{acknowledgments}

We thank  David Politzer for important contributions to this work. 
The work of BG, DO  and
MBW was supported in part by the US Department of Energy under contracts
DE-FG03-97ER40546,
 DE-FG02-90ER40542 and
DE-FG03-92ER40701,
respectively. 
\end{acknowledgments}



\end{document}